# Robust self-trapping of vortex beams in a saturable optical medium

Albert S. Reyna,[1]* Georges Boudebs[2], Boris A. Malomed[1#] and Cid B. de Araújo[1]

[1]*Departamento de Física, Universidade Federal de Pernambuco, 50670-901, Recife, PE, Brazil*

[2]*LUNAM Université, Université d'Angers, Laboratoire de Photonique d'Angers, EA 4464, 49045 Angers, France*

\# Permanent address: *Department of Physical Electronics, School of Electrical Engineering, Faculty of Engineering, Tel Aviv University, Tel Aviv 69978, Israel*

- Corresponding author. E-mail: areynao@yahoo.com.br

## Abstract

We report the first observation of robust self-trapping of vortex beams propagating in a uniform condensed medium featuring local saturable self-focusing nonlinearity. Optical vortices with topological charge $m=1$, that remain self-trapped over ~ 5 Rayleigh lengths, are excited in carbon disulfide using a helical light beam at 532 nm and intensities from 8 to 10 GW/cm$^2$. At larger intensities, the vortex beams lose their stability, spontaneously breaking into two fragments. Numerical simulations based on the nonlinear Schrödinger equation including the three-photon absorption and nonpolynomial saturation of the refractive nonlinearity demonstrate close agreement with the experimental findings.

PACS numbers: 42.65.Sf, 42.65.Jx

## I. INTRODUCTION

The spatiotemporal evolution of light beams in nonlinear (NL) media is a subject of broad interest in fundamental and applied research [1]. In transparent condensed (solid or liquid) materials, the beam propagation is generically dominated by the nonresonant Kerr nonlinearity, which induces changes in the materials' refractive index that may lead to the beam's self-focusing (or defocusing), spectral broadening, and other NL phenomena [2]. The beam propagation in centro-symmetric materials with the nonlinearity described by the third-order susceptibility, $\chi^{(3)}$, is usually modeled by the cubic NL Schrödinger equation (NLSE) [2]. Of particular interest are beams representing spatial solitons, with diverse applications to photonics, optical computing, telecommunications, etc. It is commonly known that self-focusing media allow the stable propagation of one-dimensional [(1+1)D] spatial solitons, due to the balance between the linear diffraction and self-focusing [3]. However, two-dimensional [(2+1)D] optical solitons in media with the instantaneous cubic nonlinearity are unstable, due to the catastrophic self-focusing (*critical collapse*) at high powers [4]. Nevertheless, saturation of the nonlinearity may prevent the collapse, securing stable soliton propagation. In particular, the analysis has shown that the NLSE produces stable solutions for materials exhibiting an interplay of the focusing third-order and defocusing fifth-order susceptibilities, $\text{Re}\,\chi^{(3)} > 0$ and $\text{Re}\,\chi^{(5)} < 0$, in one, two and three dimensions [5,6]. Recently, the stable propagation of (2+1)D spatial solitons in carbon disulfide, $CS_2$, supported by this mechanism, has been demonstrated experimentally [7]. On the other hand, using resonant nonlinearity in the rarefied gas of three-level atoms, which includes competing cubic and quintic nonlinearities, along with the four-wave mixing (FWM), it was possible to demonstrate the stabilization, on a long propagation distance (~20 diffraction lengths), of various soliton species, including fundamental, dipole, and vortex ones. The FWM in a nonresonant medium (glass) was exploited too to arrest the collapse of (2+1)D quasisolitons [8,9]. Furthermore, applying a nonlinearity-management procedure [10], it was possible to observe stable (2+1)D spatial solitons in a composite with suppressed $\chi^{(3)}$ but conspicuous focusing $\chi^{(5)}$ and defocusing $\chi^{(7)}$ susceptibilities [11].

In defocusing media, spatial solitons appear as optical vortices and dark solitons [12]. The vortices are axisymmetric beams with a phase singularity and zero amplitude at the pivot [13]. These helical beams carry the phase factor, $\exp(im\theta)$, where $\theta$ is the azimuthal coordinate and $m$ is the topological charge. Contrary to bright (fundamental) spatial solitons, delocalized (dark) optical vortex solitons (DOVSs), supported by a finite background, are stable structures in defocusing NL media [14]. Experimental observations of DOVSs in defocusing media were reported by several groups [14, 15]. However, bright (self-trapped) optical vortex solitons in self-focusing media are subject to spontaneous azimuthal symmetry breaking due to the corresponding modulational instability [16-20]. Many works aimed to identify suitable conditions for the stabilization of self-trapped optical vortex solitons [21-26]. In particular, bright optical vortex solitons in media combining cubic focusing and quintic defocusing nonlinearities have regions of stability and azimuthal instability, depending on the beams' power [25-27]. While this subject has been elaborated theoretically, no experimental report showing the stable propagation of a self-trapped vortex beam in a self-focusing uniform medium with local nonlinearity has been presented, thus far.

This work aims to report the first observation of effectively stable propagation of (2+1)D self-trapped vortex beams, with topological charge $m=1$, in a condensed optical medium, *viz.*, liquid $CS_2$, which features strong self-focusing [28]. The stable propagation of self-trapped vortex beams, which keep their shape and size unaltered over ~5 Rayleigh lengths, is reported here, exploiting a combination of the saturation of the refractive nonlinearity and three-photon absorption (3PA). The behavior of the self-trapped vortex beam is reproduced by using a modified NLSE which very well models the filamentation of light in $CS_2$, generated by a picoseconds laser input at 532 nm [29]. In the instability regime, splitting of the vortex beam into two separating fragments is observed at large intensities, in agreement with the numerical simulations.

## II.  THE EXPERIMENTAL SETUP

The setup used to study the vortex-beam propagation is displayed in Fig. 1. The second-harmonic beam at 532 nm, obtained from a Nd: YAG laser (80 ps, 10 Hz, 1064 nm), with the maximum pulse energy of 10 µJ, was used. An optical vortex beam with topological charge $m=1$ was produced by passing the Gaussian beam through a phase plate (VPP) manufactured by RPC Photonics. The control of the incident-beam's power was provided by a $\lambda/2$ plate followed by a Glan prism, which assures that the beam is linearly polarized. A telescope was used to adjust the beam waist, in order to illuminate a large area of the VPP, and a spatial filter was used to eliminate higher-order diffracted light. The vortex beam was focused by a 5 cm focal distance lens (L1) on the input face of a glass cell filled by $CS_2$. The waist of the Gaussian-shaped beam was 11 µm, and the vortex' core radius at the focus was 3 µm. Fig. 1(b) shows the transverse intensity profile of the beam at the input face of the sample. To confirm the presence of the topological charge carried by the beam, the triangle aperture method was used [30]. The respective diffraction pattern is shown in Fig. 1(c), where the two bright points on each side of a triangular lattice correspond to $m=1$.

Transverse vortex-beam profiles were recorded using a CCD camera aligned with the beam-propagation direction (the $z$ axis). Cells of thickness 1, 2, 3, 4 and 5 mm filled by $CS_2$ were used to image the propagation of the vortex beam over different distances, as in [31]. Lens L2 was used to obtain the beam's image at the output face with magnification $M = 4$. The imaging system, consisting of lens L2 and the CCD, can scan along $z$ to image the input and output face of the five cells, maintaining the same magnification. Small marks on the input and output faces of the cells help to identify the correct position of the imaging system, by observing a sharp image of the mark in the CCD. To observe the evolution of the vortex beam in the transverse plane, measurements were first performed with a 1 mm long cell. The imaging system was translated along the $z$ axis to image the entry (at $z = 0$) and output of the cell (at $z = 1$ mm). Then, the 1 mm long cell was replaced by a 2 mm long cell maintaining the same position of the input face, with respect to lens L1, and translating the imaging system over $z = 2$ mm. The initial position of the cell was corroborated using side-view measurements (with precision of ~1 µm), as described below. The same procedure was performed for the other cells with different thicknesses. In addition, beam images were obtained using the scattered-light imaging method (SLIM) [32], by measuring the weak scattered light in the direction perpendicular to the beam's pathway. A cell 10 mm long was used for these measurements. The setup collecting the scattered light consisted of two cylindrical lenses with 40 mm (y-axis) and 80 mm (z-axis) focal lengths, used to obtain images with magnification of 7 and $1/2$ respectively. The experiments were performed with intensities adjusted from $I = 0.5$ GW/cm$^2$ to $I = 25$ GW/cm$^2$, to identify regions of stable and unstable propagation of the vortex beam. The margin of error in the experimental measurements is given by the camera pixel size (4.6 μm) divided by the magnification. To ensure that the images correspond to the same laser pulses, both CCD cameras were triggered by Nd:YAG laser pulses, at the repetition rate of 10 Hz. Additionally, to keep control over intensity fluctuations of the laser, a post-filtering selection was carried out to keep records solely of images corresponding to the intensities varying at most by ±2%.

### III. RESULTS AND DISCUSSION

Figure 2 presents the beam profiles at the entrance and exit faces of each cell used, for two values of the laser intensity. Figure 2(a), corresponding to relatively low intensity, $I = 1$ GW/cm$^2$, shows that the vortex beam diverges along the propagation pathway without changing its ring-like shape, NL effects being negligible in this case. On the other hand, it is observed in Fig. 2(b) that, for $I = 9\,\text{GW}/\text{cm}^2$, the beams' shape and radius remain constant for the propagation over 3 mm, which corresponds to ~5 Rayleigh lengths; this result clearly indicates the formation of a stable self-trapped vortex beam. At $z > 3\,\text{mm}$, the beam diverges because the intensity is depleted by the NL absorption. As show below, numerical simulations corroborate that a long distance of the stable propagation of self-trapped vortex beams can be attained. Figures 2(c) and 2(d) show the intensity distribution along the radial coordinate corresponding to Figs. 2(a) and 2(b), respectively. The solid and dashed lines represent the evolution of the beam size, as produced from the theoretical model described below.

Figure 3 displays side-view images recorded for intensities from $1\,\text{GW}/\text{cm}^2$ to $18\,\text{GW}/\text{cm}^2$. Figures 3(a) and 3(b), in conjunction with Fig. 2(a), demonstrate that, for $I \leq 5 \, \text{GW}/\text{cm}^2$, the vortex beam does not change its ring shape, while diverging due to the linear diffraction. For $5 \, \text{GW}/\text{cm}^2 \leq I < 8 \, \text{GW}/\text{cm}^2$, the beam's divergence weakens with the increase of the intensity, due to the self-focusing effect. Figures 3(c)-3(e), obtained for $8 \, \text{GW}/\text{cm}^2 \leq I \leq 10 \, \text{GW}/\text{cm}^2$, exhibit the stable propagation of the vortex beam up to the distance of ~3 mm. Thus, Figs. 2(b) and 3(c)-3(e) provide the direct evidence for the propagation of a *stable ring-shaped vortex*. However, at $I > 10 \, \text{GW}/\text{cm}^2$, strong concentration of the power was observed in the course of the first 3 mm of the propagation, and the transverse images exhibit distortion of the beam profiles. These asymmetries gradually increase, up to splitting of the vortex beam observed at $I \geq 18 \, \text{GW}/\text{cm}^2$, as shown in Fig. 3(f). The low resolution of the image after the splitting is due to the weakness of the scattered light. Figures 3(g)-3(l) present the variation of the beam's radius in the course of the propagation, corresponding to Figs. 3(a)-3(f), respectively. Shaded rectangles display the region where the self-trapped vortex beam is *stable*.

In the range of intensities used, the NLSE for CS$_2$ has to be modified to include additional effects, which depend on the wavelength and pulse duration [28]. In particular, the combination of $\chi^{(3)}$ and $\chi^{(5)}$ terms was used to explain the formation of bright spatial solitons excited by 100 fs laser pulses at 920 nm [7]. On the other hand, illuminating CS$_2$ by 12 ps pulsed beams at 532 nm, it was concluded in Ref. [29] that, for the range of intensities tested, the nonlinearity is properly described by a nonpolynomial refractive index, $n_{2,\text{eff}}(I) = aI/(1+b^2 I^2)$, with $a = 6.3 \times 10^{-33} \, m^4/W^2$ and $b = 2.3 \times 10^{-15} \, m^2/W$. This specific expression (*ansatz*) for the NL refractive index was adopted to fit data obtained for CS$_2$ in the ps regime by means of the D4σ method [33], which provides very accurate measurements for the second moment of the intensity distribution of the transmitted beam. Instead of measuring the variation of the transmitted intensity, as the case of the Z-scan technique, the D4σ method directly measures changes in the transverse profile of the transmitted beam, thus leading to a more exact value of the NL refractive index [33]. Experimental data reported in Ref. [29] clearly show saturable-refraction behavior of the $CS_2$ medium for intensities on the order of tens of GW/cm$^2$. On the other hand, the usual expression for the saturable nonlinearity, $n_{2,\text{eff}}(I) = a'/(1+b'I)$, adopted in Ref. [16], does not adequately describe the experimental results for low intensities. Figure 4(a) shows a comparison between the two NL-refractive *ansatz* proposed in Refs. [16] and [29], where values of $a' = 20 \times 10^{-18}$ m$^2$/W and $b' = 19.3 \times 10^{-15}$ m$^2$/W are used to compare the results corresponding to the model of Ref. [16] (the red line) with their counterparts reported in Ref. [29] (the black line). Note that both models produce similar behavior of $n_{2,\text{eff}}$ for high intensities. Thus, the saturation *ansatz* from Ref. [29] is expected to produce the instability of the vortex beam, leading to splitting of the vortex, similar to results of Ref. [16]. However, for low intensities the model from Ref. [16] does not describe the experimentally observed stability region of the self-trapped vortex beams in $CS_2$ [the shaded vertical rectangle in Fig. 4(a)].

However, it should be noted that the model proposed in Ref. [29] is no more valid for low intensities (the region corresponding to the dashed line). The measurements reported in Ref. [29] show that $n_2 \approx 1.4 \times 10^{-14}$ cm$^2$/W for intensities in the range of 1-2 GW/cm$^2$ [34]. Indeed, the signal obtained by using the D4σ method becomes very weak at this level, therefore the growth of $n_{2,\text{eff}}$ between $1.4 \times 10^{-14}$ cm$^2$/W and $\approx 3 \times 10^{-14}$ cm$^2$/W at intensity 20 GW/cm$^2$ may be understood as a contribution of the self-focusing fifth-order nonlinearity. At larger intensities, the effect of the plasma generation, characterized by the NL absorption, makes the self-defocusing NL refraction more and more dominant, contributing to the reduction of $n_{2,\text{eff}}$. This behavior can be understood as a physical explanation of the overall variation observed in Fig 4(a).

As concerns the NL absorption, it was also concluded in Ref. [29, 34] that the two-photon absorption is negligible, while the three-photon absorption (3PA) must be taken into account, with respective coefficient $\gamma = 9.3 \times 10^{-26}$ m$^3$/W$^2$. Figure 4(b) shows the intensity transmitted through the CS$_2$ sample (the cell length being 1 mm) versus the incident-beam's intensity, for the vortex beam with $m = 1$. Blue circles represent experimental data collected under the same conditions which were used to study the propagation of the self-trapped vortex beam. The solid line represents the theoretically calculated evolution of the optical intensity, $I$, along the propagation distance, $z$, produced by the respective differential equation, $dI(z)/dz = -\gamma I^3(z)$, which implies that the absorption in CS$_2$ is determined solely by the 3PA, as concluded in Ref. [29, 34]. It is seen that, with the value of $\gamma$ reported in Ref. [29], the experimental and theoretical results are in very good agreement.

To describe the propagation of the vortex beams in CS$_2$ in the picosecond regime, we used a modified NLSE, which includes the saturable refractive index, and the 3PA as per Ref. [29]:

$$i\frac{\partial E}{\partial z} = -\frac{1}{2n_0 k}\Delta_\perp E - \left(\frac{kaI^2}{1+b^2 I^2} + i\frac{\gamma I^2}{2}\right)E, \quad (1)$$

where $E$ is the field amplitude $\left(I = 2\varepsilon_0 n_0 c |E|^2\right)$, $\Delta_\perp$ the transverse Laplacian, $z$ the propagation distance, $k = 2\pi/\lambda$ ($\lambda$ is the carrier wavelength), $n_0$ the linear refractive index, and $b$ the saturation coefficient. We normalize the variables as $X = x/w_0$, $Y = y/w_0$, $Z = z/L$, $U = E/E_r$, with $L = n_0 k w_0^2$ and $E_r = (2bn_0 c\varepsilon_0)^{-1/2}$, where $w_0$ is the initial beam's waist, $c$ the speed of light in vacuum, and $\varepsilon_0$ the vacuum permittivity, to rescale Eq. (1):

$$\frac{\partial U}{\partial Z} = \frac{i}{2}\left(\frac{\partial^2 U}{\partial X^2} + \frac{\partial^2 U}{\partial Y^2}\right) + i\frac{\eta |U|^4 U}{1+|U|^4} - \mu |U|^4 U, \tag{2}$$

where $\eta \equiv Lka/b^2$ and $\mu \equiv \gamma L/(2b^2)$. The intensity is related to the normalized field by $I = |U|^2/b$.

Simulations of Eq. (2) were initiated with the input wave form $U(R,\theta,Z=0) \propto \exp(-R^2/w_0^2 + im\theta)\tanh[R/(2w_v)]$, where $R$ and $\theta$ are the polar coordinates and $m$ is the topological charge, $w_0$ and $w_v$ being waists of the Gaussian background and vortex core, respectively. Numerical results for the vortex-beam propagation in the 10 mm long cell filled by CS$_2$ were produced for $L = 2.3$ mm, $\eta = 28$ and $\mu = 3.3$.

Figure 5 shows the evolution of the transverse beam's profiles for intensities between $1\,\text{GW/cm}^2$ and $15\,\text{GW/cm}^2$, obtained by simulations of Eq. (2), which were performed by means of the split-step compact finite-difference method [35]. Figure 5(a) displays the divergence of the vortex beam for $I = 1\,\text{GW/cm}^2$, which is similar to what happens in the linear regime, according to Figs. 2(a) and 3(a). For $I = 9\,\text{GW/cm}^2$, the propagation of the self-trapped vortex beam can be observed for over distance ~3 mm, as shown in Fig. 5(b), which accords with Figs. 2(b) and 3(d). Figure 5(c) shows a deformation of the beam's profile for $I = 12\,\text{GW/cm}^2$, which gradually grows, leading to the complete split of the vortex beam at $I = 15\,\text{GW/cm}^2$, as shown in Fig. 5(d).

Figure 6 shows a longitudinal cross section of the vortex-beam propagation, produced by simulations of Eq. (2). At $I = 1\,\text{GW}/\text{cm}^2$ [Fig. 6(a)], the vortex beam keeps the ring shape but diverges due to the diffraction. On the other hand, in Fig. 6(b), corresponding to $I = 8.5\,\text{GW}/\text{cm}^2$, the beam slightly diverges at first, but, after passing ~1.2 mm, it keeps constant shape and width in the course of the propagation over ~3 mm, and diverges afterwards. Figure 6(c) shows the variation of the vortex-beam radius at several positions in the cell for different intensities, the shaded rectangles displaying the region of the *stable propagation* of the (2+1)D self-trapped vortex beams, for $I = 8\,\text{GW/cm}^2$ and $9\,\text{GW}/\text{cm}^2$. For $15\,\text{GW}/\text{cm}^2$, the curve ends at $z$ = 3 mm, as the vortex splits in two fragments beyond this point.

To highlight the effect of the 3PA ($\gamma > 0$), Fig. 6(d) shows the evolution of the vortex-beam radius produced by simulations of Eq. (2) with $\gamma = 0$. In this case, at $I < 7\,\text{GW}/\text{cm}^2$ the beam diverges, like in Fig. 6(c). At $I = 7.5\,\text{GW}/\text{cm}^2$, it initially diverges, passing 1.2 mm, but features stable propagation of the self-trapped vortex beam over subsequent 2.5 mm. For $I$ = 8 GW/cm$^2$, the vortex is unstable, splitting in two fragments.

Figure 7(a) shows, at $I = 18\,\text{GW}/\text{cm}^2$, two bright fragments of radius 17 μm at the output face, with distance 68 μm between them, the intensity of each one being 10% of the initial value. The intensity loss is caused by the 3PA, while the difference between the fragments results from a small asymmetry in the input beam. Figure 7(b) shows the respective numerical result, obtained from Eq. (2) for $I = 15\,\text{GW}/\text{cm}^2$. The spiral emerging around the fragments in the simulations (it is more salient at $I \geq 16\,\text{GW/cm}^2$) was not observed in the experiment, as the camera was not sensitive enough for that.

The experiment was repeated for the input beam with vorticity $m=-1$, obtained by reversing the input face of the VPP. Figures 7(c, d) display the respective experimental and numerical results, with two fragments similar to those in Figs. 7(a, b), but rotated by 90°. The experimental and related numerical images obtained for $m=-1$ demonstrate that the results are highly reproducible. Similar results have been obtained for other input intensities.

In the simulations, the fragments emerging after the splitting of the vortex beam move along tangents to the vortex ring, due to conservation of the orbital angular momentum (see Ref. [36]). However, unlike previous theoretical results which predict the formation of fundamental solitons after the splitting [16, 17], in the present case the fragments are not solitons, because of the losses induced by the 3PA. The model used here can be applied to the propagation of vortex beams with multiple topological charges too, but they tend to be unstable against splitting, unlike the vortex with $m = 1$. In particular, simulations (not shown here) demonstrate that a vortex beam with $m=2$ spontaneously splits into a ring-shaped set of four bright fragments (which cannot be identified as fundamental solitons with $m = 0$), which move due to the conservation of angular momentum.

Simulations were also performed with $I/(1+b^2I^2)$ in Eq. (2) replaced by $1/(1+bI)$, which is the form of the saturable nonlinearity adopted in Ref. [16]. Varying the input intensity from 1 to 25 GW/cm², *no stability region* for self-trapped vortex beams was found in that case. Thus, the crucially important ingredients necessary for the stable propagation of the self-trapped vortex beams are the appropriate intensity dependence of the NL refractive index, as derived in Ref. [29], and the 3PA. Actually, the 3PA term in Eq. (2) helps to *expand* the stability region for self-trapped vortex beam. In particular, with this term kept in Eq. (2), the splitting of the vortex in two fragments is observed at $I > 13\,\text{GW}/\text{cm}^2$, while the stable propagation occurs at $8\,\text{GW/cm}^2 \leq I < 10\,\text{GW/cm}^2$. However, if the 3PA term is dropped, the splitting occurs at $I \geq 8\,\text{GW}/\text{cm}^2$, with a tiny stability region spotted at $7.4\,\text{GW}/\text{cm}^2 \leq I \leq 7.6\,\text{GW}/\text{cm}^2$.

## IV. SUMMARY

In summary, for the first time the observation and characterization of (2+1)D self-trapped vortex beams, which stably pass ~ 5 Rayleigh lengths, is reported using a condensed medium (liquid $CS_2$) with the local saturable self-focusing nonlinearity. The self-trapped vortex beams with topological charge $m=1$ are *azimuthally stable* at moderate values of the input intensity, due to the saturation of the refractive nonlinearity and the instability-suppressing effect of the 3PA (three-photon absorption). At higher intensities, the vortex beams are unstable, spontaneously splitting into a pair of separating fragments. The experimental findings are accurately modeled by the modified NLSE with the saturable NL refractive index and the 3PA coefficient gathered from recent measurements [29].

Strictly speaking, the stability of the self-trapped vortex beams reported here is a transient effect, as the 3PA eventually causes degeneration into the linear regime, while the saturable refractive nonlinearity alone cannot stabilize self-trapped vortex beams in the absence of the nonlinear loss [1, 16]. On the other hand, as mentioned above, the analysis predicts a stability region for self-trapped (bright) optical vortex solitons in the conservative medium with the cubic-quintic (rather than saturable) nonlinearity [24-27]. Recently, this nonlinearity was implemented in colloidal samples, the loss being negligible at experimentally relevant propagation distances [10, 11]. The work aimed at the creation of unconditionally stable bright optical vortex solitons in this setting is currently in progress.

## ACKNOWLEDGMENTS

This work was supported by Brazilian agencies Conselho Nacional de Desenvolvimento Científico e Tecnológico (CNPq) and the Fundação de Amparo à Ciência e Tecnologia do Estado de Pernambuco (FACEPE). The work was performed in the framework of the National Institute of Photonics (INCT de Fotônica - INFO) project and PRONEX/CNPq/FACEPE. We appreciate valuable discussions with A. M. Amaral, E. L. Falcão-Filho, Y. V. Kartashov, and L. Torner. The donation of the vortex phase plate by RPC Photonics is also acknowledged. G.B. and B.A.M. acknowledge hospitality of the Department of Physics at Universidade Federal de Pernambuco.

**Figure captions**

1. (Color online) (a) The experimental setup: polarizer (P); mirror (M); telescope (T); vortex phase plate (VPP); spatial filter (SF); spherical lenses with $f_1 = 5\,\text{mm}$ (L1) and $f_2 = 5\,\text{mm}$ (L2). The CCD1 camera produced the transmitted-beam spatial profile. Cylindrical lenses with $f = 40\,\text{mm}$ (CL1) and $f = 80\,\text{mm}$ (CL2), and CCD2 were used in the SLIM setup. The cell's length is 10 mm. (b) The intensity profile of the input vortex beam. (c) The diffraction pattern of the beam with topological charge *m* = 1, produced by the triangle aperture method.

2. (Color online) Transverse vortex-beam profiles at input and output faces for cells with thicknesses 1, 2, 3, 4 and 5 mm: (a) $I = 1\,\text{GW}/\text{cm}^2$ and (b) $I = 9\,\text{GW}/\text{cm}^2$. The lines are guides to the eye. (c)–(d) Normalized intensity distributions of the beam at each position from (a)–(b).

3. (Color online) Experimental side-view images of the vortex-beam propagation for intensities (a) $1\,\text{GW}/\text{cm}^2$, (b) $5\,\text{GW}/\text{cm}^2$, (c) $8\,\text{GW}/\text{cm}^2$, (d) $9\,\text{GW}/\text{cm}^2$, (e) $10\,\text{GW}/\text{cm}^2$ and (f) $18\,\text{GW}/\text{cm}^2$. (g)-(l) The beam's radius as a function of the propagation distance, corresponding to (a)-(f), respectively. The shaded areas indicate the region of the *stable vortex-beam propagation*.

4. (Color online) (a) The effective NL refractive index of $CS_2$ as a function of the laser intensity. Red and black lines correspond to the models adopted in Refs. [16] and [29], respectively. The vertical blue bar indicates the intensity range of the effective stability of the self-trapped vortex beams. (b) Transmittance of $CS_2$ versus the input laser intensity, in the 1 mm thick cell. The solid line corresponds to the theoretical fit corresponding to the 3PA effect.

5. (Color online) Numerically generated images showing the evolution of transverse vortex-beam profiles (with m = 1) along the propagation direction, for intensities $1\,\text{GW}/\text{cm}^2$ (a), $9\,\text{GW}/\text{cm}^2$ (b), $12\,\text{GW}/\text{cm}^2$ (c), and $15\,\text{GW}/\text{cm}^2$ (d).

6. (Color online) Numerical results for the vortex-beam propagation at (a) $I=1\,\text{GW}/\text{cm}^2$ and (b) $I=8.5\,\text{GW}/\text{cm}^2$, obtained by simulations of Eq. (2). (c, d) The beam's radius as a function of the propagation distance, produced by the simulations for the 3PA coefficient (c) $\gamma=9.3\times10^{-26}\,\text{m}^3/\text{W}^2$ and (d) $\gamma=0$, with various intensities. The shaded area in (c) indicates the intensity range of stable propagation of self-trapped vortex beams.

7. (Color online) (a, c) Experimental images obtained in the output face of the cell, after the splitting of the vortex beam with topological charge $m=+1$ (a) and $m=-1$ (c), for laser intensity $I=18\,\text{GW}/\text{cm}^2$. (b, d) Simulations of Eq. (2) for (b) $m=+1$ and (d) $m=-1$, at $I=15\,\text{GW}/\text{cm}^2$.

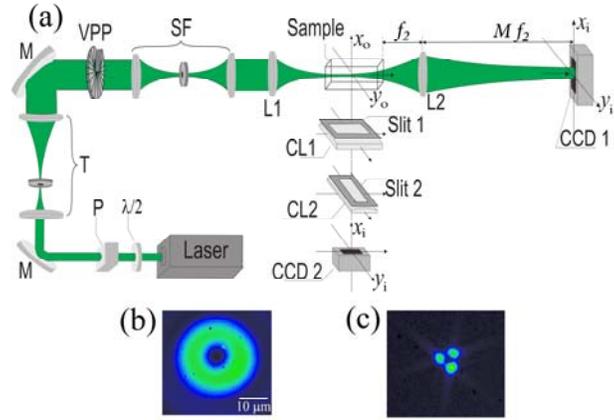

Fig. 1 Reyna et al.

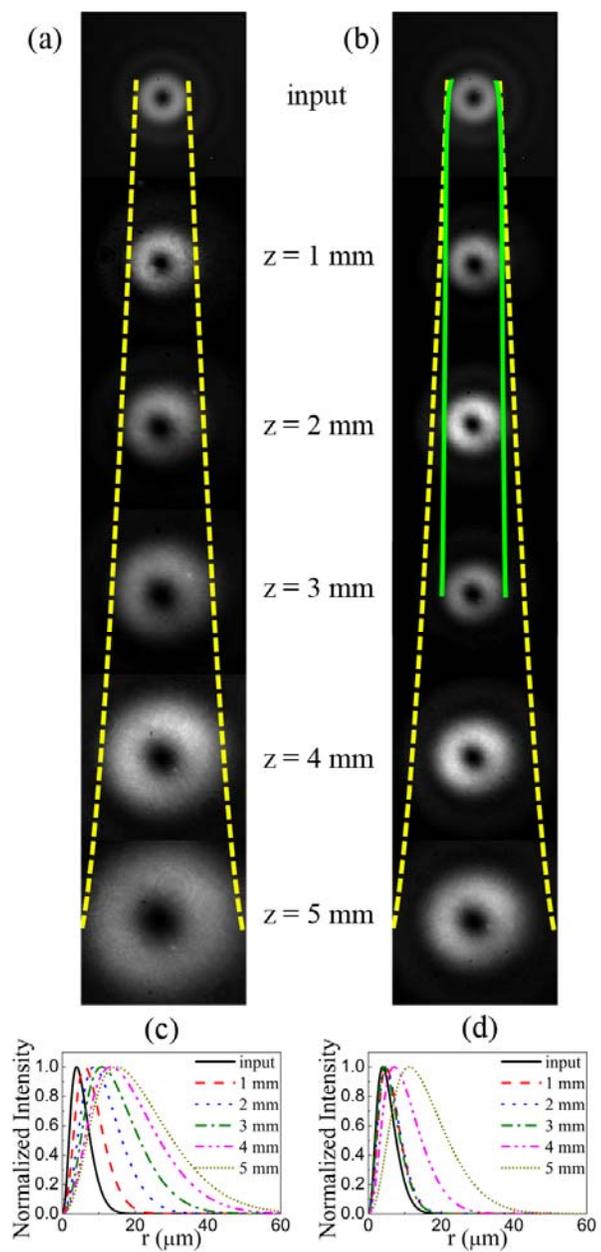

Fig. 2 Reyna et al.

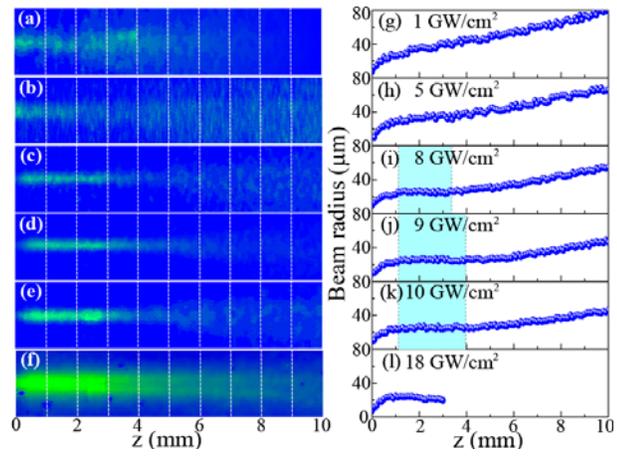

Fig. 3 Reyna et al.

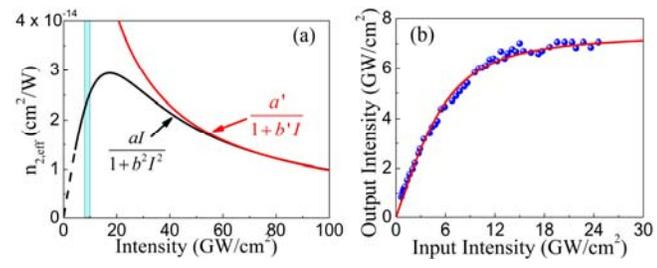

Fig. 4 Reyna et al.

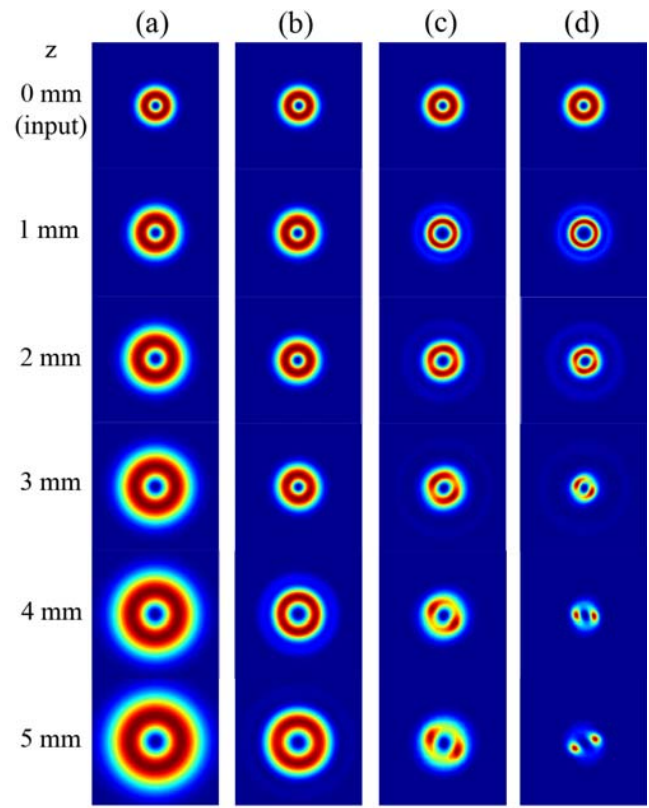

Fig. 5 Reyna et al.

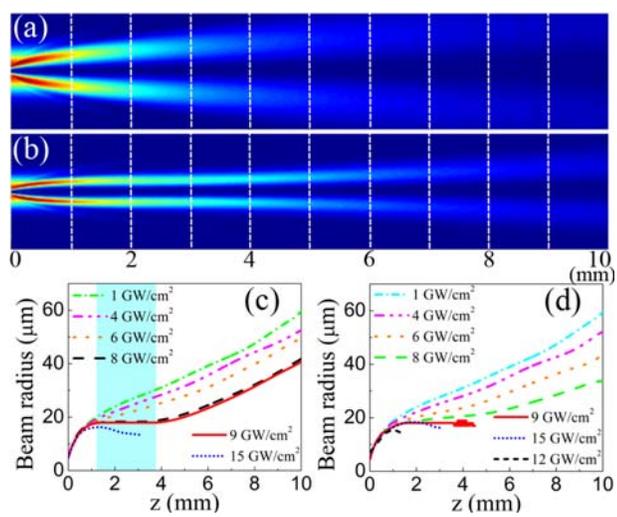

Fig. 6 Reyna et al.

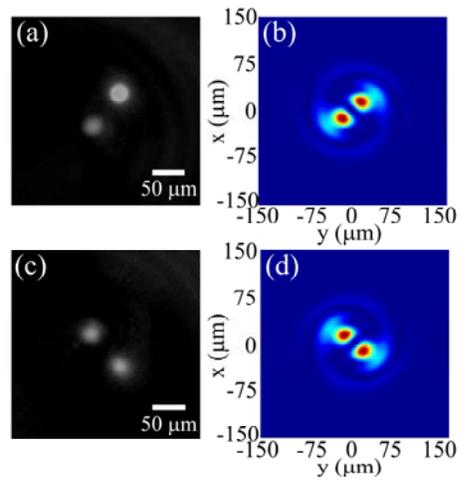

Fig. 7 Reyna et al.